\documentstyle[aps]{revtex}

\begin{document}
\author{Liu Yaowen,\thanks{%
E-mail address: chaosun@lzu.edu.cn} Li Haibin, Zhao Hong,\thanks{%
E-mail address: zhaoh@lzu.edu.cn} Wang Yinghai\thanks{%
Author to whom correspondence should be addressed. E-mail: wangyh@lzu.edu.cn}}
\address{Department of Physics, Lanzhou University, Lanzhou 730000, China}
\title{Self-similarity in a system with a short-time delayed feedback}
\date{\today}
\maketitle

\begin{abstract}
Using the Poincar\'{e} section technique, we study in detail the dynamical
behaviors of delay differential system and find a new type of solutions $S_i$
in short-time delay feedback. Our numerical results remind us to deny the
opinion that there are no complex phenomena in short-time delay case. Many
similarities between foundamental solution and the new type of solutions are
found. We demonstrate that the scales of $S_i$ increase with exponential
growth via $i$ in the direction of $\mu $, while decrease with exponential
decays in the direction of $x$ or delay time $t_R$.
\end{abstract}


\section{Introduction}

Optical feedback systems governed by delay differential equations (DDEs)
have attracted much attention from both the applied and the fundamental
points of view [1-16]. Generally, the delay-differential system related to
optical bistable or hybrid optical bistable device is described by 
\begin{equation}
\tau ^{\prime }\dot{x}(t)=-x(t)+f(x(t-t_R),\mu ),  \label{aaa}
\end{equation}
where $x(t)$ is the dimensionless output of the system at time $t,$ $t_R$ is
the time delay of the feedback loop, $\tau ^{\prime }$ is the response time
of the nonlinear medium, the parameter $\mu $ is proportional to the
intensity of the incident light. In Eq. (\ref{aaa}), $f(x,\mu )$ is a
nonlinear function of $x$, characterizing the system, e.g. $f(x,\mu )=\mu
\pi [1-\zeta \cos (x-x_B)]$ for Ikeda model \cite{Ikeda}, $f(x,\mu )=\pi
[A-\mu \sin ^2(x-x_0)]$ for Vall\'{e}e model \cite{Vallee}, and $f(x,\mu
)=\mu \sin ^2(x-x_0)$ for the sine-square model \cite{Goedgebuer}.

The understanding of the Eq. (\ref{aaa}) up to now can be summarized as
follows. The first experimental observation of period-doubling bifurcations
and chaos in a hybrid bistable device was made by Gibbs {\it et al}. \cite
{Gibbs} following a prediction by Ikeda {\it et al}. \cite{Ikeda}. The
solution of system, which appears after Hopf bifurcation, evolves through a
period doubling $T_F\rightarrow 2T_F\cdots \rightarrow 2^NT_F$, as one
parameter is varied. These solutions are called $2^N$ periodic and the
cascade accumulates at the Feigenbaum point. These solutions are named {\it %
fundamental solutions} by Ikeda {\it et al}. \cite{Ikeda}, we do so in this
paper. Later the two groups found that {\it higher- harmonic} oscillation
states appear successively in the course of transition to developed chaos in
long-time delayed case (i.e. delay time is longer enough than the response
time). These solutions coexist and each follows a period-doubling cascade.
The oscillation period of these harmonic states are given by $T_F/n$, where $%
n$ is odd integer and $T_F$ is the period of the fundamental solution. As
the study of the dimension of the chaotic attractor, the behavior of the DDE
exhibits high-dimensional chaos \cite{Dovizes}. Ikeda and Matsumoto have
given an estimate of the Lyapunov dimension of attractor for the Ikeda
model, and it ranges approximately from $2$ to $13$ when some bifurcation
parameter is varied. Recently, some researchers demonstrated that the
behavior of quasiperiodicity followed the hierarchy of the Farey tree \cite
{Baums,Mark,Sacher,Ye,Zhao1} and the chaotic itinerancy phenomenon switches
among some different unstable local chaotic orbits \cite
{Otsuka,Fischer,Masoller}. We reported two new types of solutions found in
moderate-time and short-time delay regimes \cite{Zhao2}, which are different
from {\it the fundamental solution} and {\it the odd harmonic solution}. In
this paper, we study in detail the dynamical behaviors of the new type of
solution found in short-time delay regimes.

This paper is organized as follows: In Sec. II, the numerical methods used
in this paper are introduced. By using Poincar\'{e} section technique, we
can easily observe the course of bifurcation of DDE, and easily distinguish
the new type of solutions and the fundamental solution. In Sec. III, we
demonstrate our numerical results. In the short-time delay case, there is a
new type of solutions $S_i(i=1,2,3...),$which has many similarities
comparing with fundamental solution. Moreover, these new solutions are alike
each other. In Sec. IV we summarize our results and conclude.

\section{Numerical Methods}

Measuring the delay time in units of $t_R$, one can rewrite Eq. (\ref{aaa})
as 
\begin{equation}
\tau \dot{x}(t)=-x(t)+f(x(t-1),\mu ),  \label{model}
\end{equation}
where $\tau =\tau ^{\prime }/t_R$ characterizes the effect of the time delay
when $\tau ^{\prime }$ is fixed. In this paper, we study Eq.(\ref{model})
with the special feedback function 
\begin{equation}
f(x,\mu )=1-\mu x^2.  \label{Hao}
\end{equation}
This feedback function can be considered as the first nonlinear term of the
Taylor expansion of the general nonlinear function $f(x,\mu )$ in the
vicinity of a steady state. It should keep the general nonlinear properties
of DDE, as shown in Refs.\cite{Li,Zhao2}.

The Eq. (\ref{model}) can be solved numerically and a fourth-order Adam's
interpolation is suitable for that. In order to trace the evolution of a
DDE, one might investigate the evolution curve of the variable $x(t)$ vs the
time $t.$ However, it is difficult to distinguish different solutions if one
only observes the $x(t)-t$ relation. Some of us (Zhao {\it et al.}) have
offered a method in Ref. \cite{Zhao1} to represent the solutions of a
one-variable DDE by using the Poincar\'{e} section technique. This method
has been proved to be a powerful tool in exploring the evolution of
bifurcation of DDE. Let us review this method briefly. Let $x_t(\theta
)\equiv x(t+\theta ),-1\leq \theta \leq 0,$ then $x_{t_2}(\theta )$ is
determined by $x_{t_1}(\theta )$ uniquely according to Eq. (\ref{model}),
where $t_1<t_2$. We approach the section mapping as follows: choose an
appropriate constant $x_c\in R$; integrate Eq. (\ref{model}) numerically
till $x(t)>x_c$ and $x(t+h)<x_c,$ where $h$ is the length of the integrating
step; then proceed a simulation procedure to get $t_i$ as well as $%
x_{t_i}(\theta )$ such that $x_{t_i}(0)=x_c$. To be simple, we denote $%
x_{t_i}(\theta )$ as $x_i(\theta )$ in the following discussions. In this
way we convert the flow of Eq. (\ref{model}) into a mapping which maps the
curve $x_i(\theta )$ onto the curve $x_{i+1}(\theta )$. We regard this
curve-to-curve mapping as the Poincar\'{e} map of a DDE. A periodic solution
of Eq. (\ref{model}) with period $T$, $x(t)=x(t+T),$ corresponds to a
periodic solution of the Poincar\'{e} map with period $N$, $x_i(\theta
)=x_{i+N}(\theta ),$ where $N$ is an integer. For practical applications, we
can take $n$ discrete points $x_i(\theta _j)$ on the curve $x_i(\theta )$ to
represent the solution, where $\theta _j\in (-1,0)$ and $j=1,2,...,n$. Then
the curve-to-curve mapping appears as a point-to-point mapping in $R^n.$ In
order to exhibit the bifurcation process, here we usually need a
one-dimensional mapping representation $x_i(\theta _1)$ with the bifurcation
parameter.

\section{Results}

As usually considered, there is no complex phenomenon in short-time delay
region since Eq. (\ref{model}) will approach a normal one-dimensional
ordinary differential equation. Our results remind us that this is not the
truth.

\subsection{The bifurcation of fundamental solution}

Before discussing the new type of solutions, let us first review the
bifurcation process of the fundamental solution. In the long-time delay case
(i.e. $\tau $ is very small), Ikeda {\it et al.} \cite{Ikeda} have shown
theoretically that instabilities and chaotic behaviors can occur in the
system. As $\tau $ $\,$is fixed and $\mu $ is increased, a square-wave
solution appears after the Hopf bifurcation of a steady state. With further
increase of $\mu $, this square wave solution undergoes a square of
bifurcation with its period doubling itself successively and then becomes
chaotic. We define this solution as {\em the fundamental solution }of the
system and marked it as $S_0.$

When $\tau $ increases (i.e. delay time decreases), the fundamental solution
exhibits mirror-similar bifurcation behavior as shown in Fig. 1(a). With the
continuous increase of $\tau $, the period-doubling bifurcation with less
and less order takes place in the course of bifurcation. At $\tau =1.13,$ $%
S_0$ undergoes only period-two bifurcation via $\mu .$ Fig.1(b) shows a
bifurcation diagram just below this value. With further increase $\tau $, we
can no longer observe the period-doubling bifurcation of $S_0$, see Fig.
1(c)-(f). We regard the regime of $\tau >1.13$ as short-time delay case. In
this regime, the delay time $t_R$ is smaller than the response time $\tau
^{\prime }$ because $\tau =\tau ^{\prime }/t_R,$ and the fundamental
solution exhibits only one-period limit cycle state.

\subsection{New type of solutions}

In the short-time delay case, fundamental solution shows no bifurcation and
chaos. This is not to say that there is no chaos in this system with the
varying of $\mu $ at a fixed parameter $\tau .$ In fact, there still exists
another chaotic attractor, which locates behind $S_0$ in the direction of $%
\mu ,$ as shown in Fig. 1(c)-(f). We marked them as $S_i(i=1,2,3,...).$ In
the direction of $\mu ,$every $S_i$ evolves a period-doubling bifurcation.
Figure 2 exhibits the evolution courses of fundamental solution and $S_i,$%
which is located on period-one limit cycle state of themselves, respectively.

From Fig. 1 and Fig. 2, we can easily find that there are not only
similarities but also difference among $S_0$ and $S_i.$ Firstly, with the
increase of $\tau ,$ each $S_i$ appears the same bifurcation process as $S_0$
does. Comparing Fig. 1(c) with Fig. 1(a), we can find the diagram of $S_1$
at $\tau =3.0,$ displays the same shape as that of $S_0$ at $\tau =0.80.$ In
fact, at certain parameter, $S_2$, $S_3$ and $S_4$ {\it et al}. also show
the similar shape. Secondly, $S_i$ has more and more oscillation with the
increase of the subscript $i.$ Simultaneously, $i$ increases with the
increase of $\tau .$ $S_0$ has only one peak within one period. In contract
to $S_0,$ $S_1$ has not only the peak but also another small peak; $S_2,S_3$
and $S_4$ have more and more peaks within one period respectively, see Fig.
2. Thirdly, with the increase of $\tau ,$ $S_i$ appears one by one and $S_1$
follows $S_0$, $S_2$ follows $S_1$ in the direction of $\mu .$

\subsection{The scales of $S_i$ via $i$}

From Fig. 1 one can find that $S_i$ appears continually with the increase of 
$\tau $. In order to find the law which exhibits the appearance order of $%
S_i $ and the scales of $S_i$, we should choose a standard to compare $S_i$
with each other. In this paper, we choose the critical values of $\tau _i$
as the standard, $\tau _i$ is the value at which the second period-doubling
bifurcation of the period-1 solution of $S_i$ takes place with the decrease
of $\tau .$ In fact one could also choose another standard. For this
specific choice, the bifurcation diagram of $S_i$ appears as the patterns in
Fig. 3(a)-(e) respectively. Our numerical solutions show that $\tau _i$
increases with exponential growth against the increase of $i.$ Figure 4(a)
demonstrates the result, where the scatters are the values of $\tau _i$ and
the dotted line is the fitting curve which is exponential growth function
with the increase of $i$ as follows: $\tau _i=A\exp (i/B),$ where $A$ and $B$
are fitting coefficients..

From Fig. 1 one can find the scale of the bifurcation diagrams of $S_i$
along $\mu $ increases, while along $x$ decreases with the increase of $i$.
Using the standard chosen above, we can measure the length $\delta \mu $ of
the period-2 solution in the direction of $\mu $ at $\tau _i$ and use it to
characterize the scale of $S_k$ along $\mu $. Figure 4(c) shows that $\delta
\mu $ also increases with exponential growth via $i$, and the exponential
function is $\delta \mu \left( i\right) =A\exp (i/B).$ On the other hand, by
measuring the maximum height $\delta x$ of the period-2 solution at $\tau
_i, $ we find that $\delta x$ exhibits the exponential decay with the
increase of $i$: $\delta x(i)=A\exp (-i/B),$see Fig. 4(d).

\section{Conclusion}

In this paper we have studied in detail the dynamics of short-time delay
differential system. Our numerical result remind us that it is not truth to
consider that there is no complex behavior in short-time delay feedback. By
using the Poincar\'{e} section technique in DDE \cite{Zhao1}, a new type of
solutions $S_i$ was found in this case and it has many similarities as same
as fundamental solution in the bifurcation diagrams. We found the law of $%
S_i $ and showed the scales of $S_i$ with the increase of $i.$ The scales of 
$S_i $ increases along $\mu $ and $\tau $, while decreases along $x$ and
delay time .

\begin{center}
{\bf ACKNOWLEDGMENT}
\end{center}

This work is supported in part by the National Natural Science Foundation of
China, and in part by Doctoral Education Foundation of National Education
Committee.

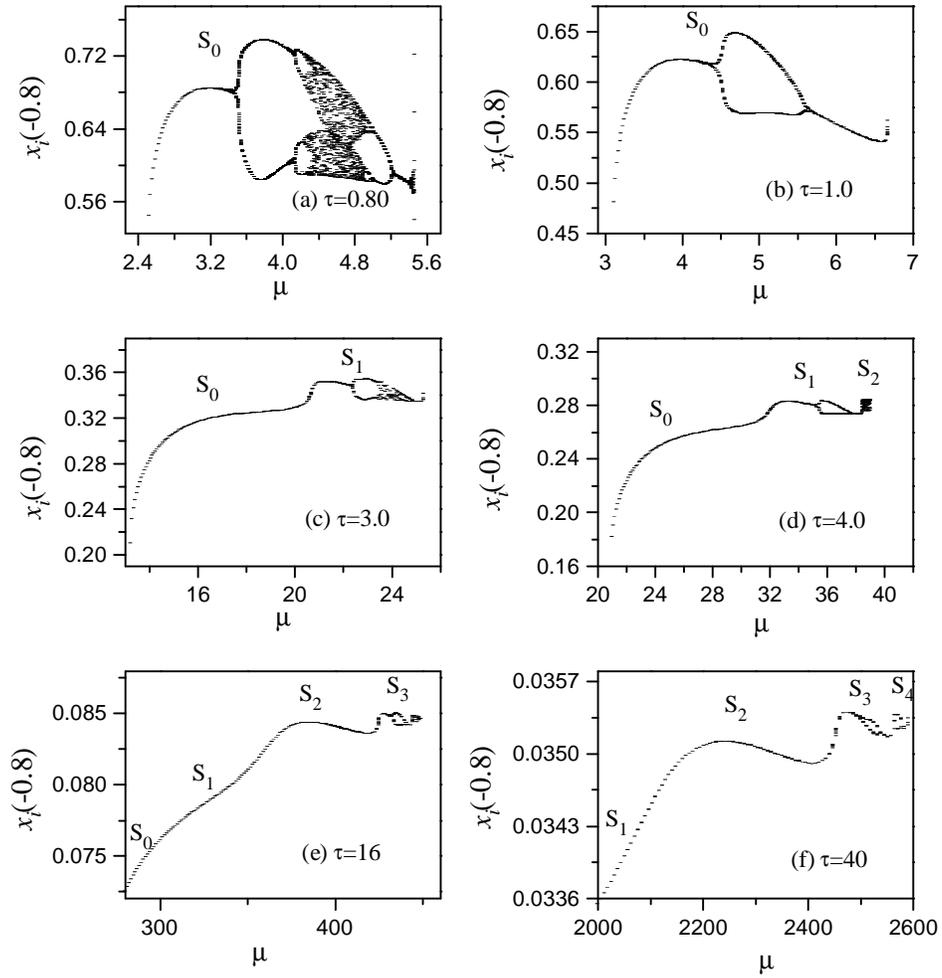
\begin{figure}[]
\caption{The bifurcation diagrams of $S_0$ and $S_i$ with the increase of $%
\tau .$}
\label{fig1}
\end{figure}

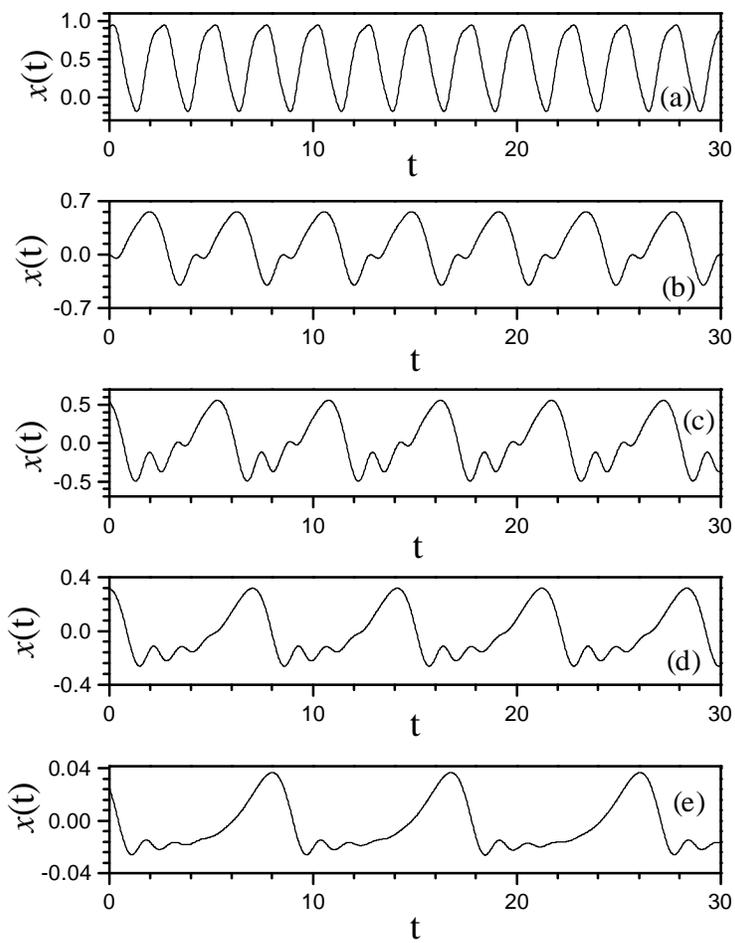
\begin{figure}[h]
\caption{The evolution of $S_0$ and $S_i$ versus time. (a) $S_0,\tau
=0.3,\mu =1.50;$ (b) $S_1,\tau =1.5,\mu =8.50;$ (c) $S_2,\tau =1.5,\mu
=9.42; $ (d) $S_3,\tau =3.5,\mu =31.45;$ (b) $S_4,\tau =40,\mu =2585. $}
\label{fig2}
\end{figure}

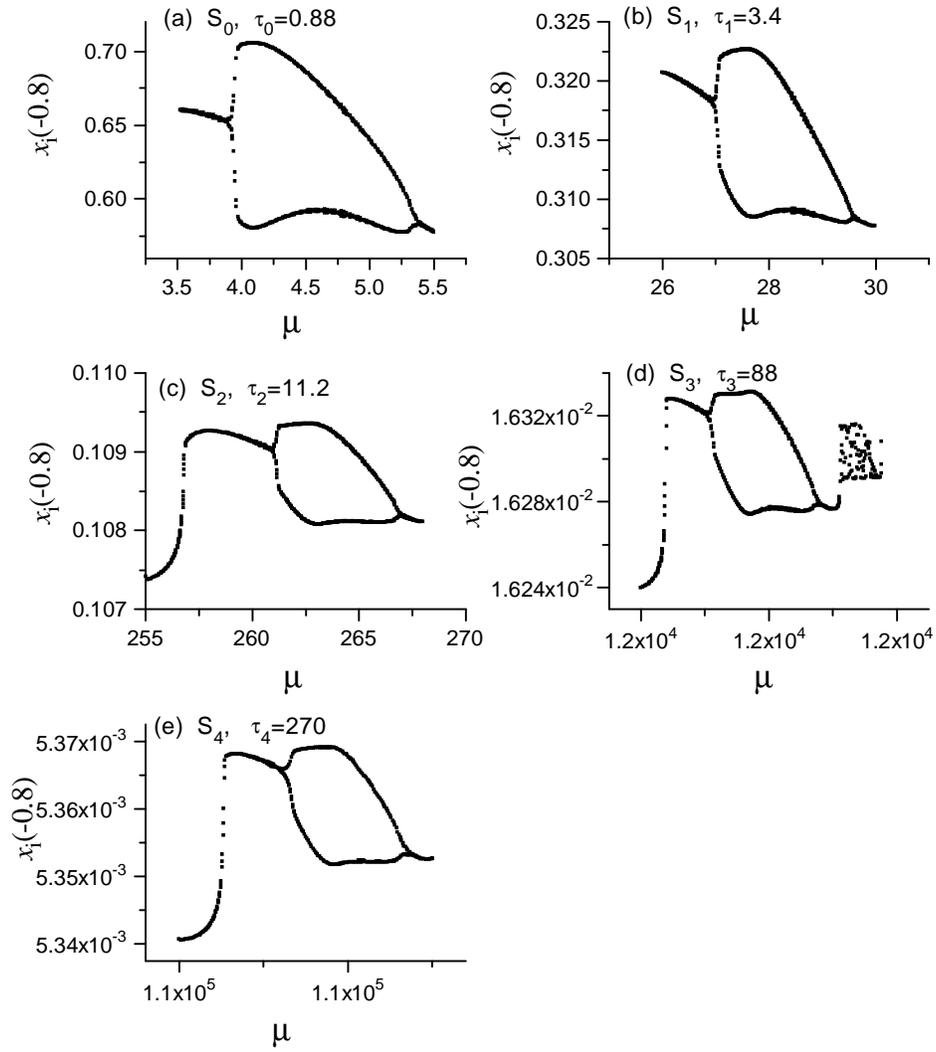
\begin{figure}[h]
\caption{Bifurcation diagrams of $S_0$ and $S_i$ at the critical values of $%
\tau _i.$}
\label{fig3}
\end{figure}

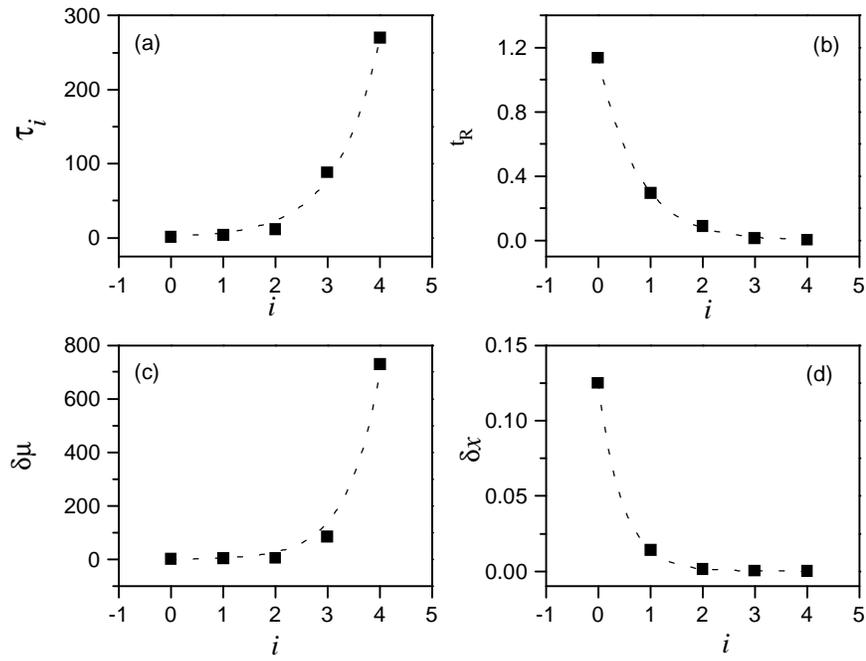
\begin{figure}[tbp]
\caption{(a) $\tau _i$, (b) $t_R$, (c) $\delta \mu $, and (d) $\delta x$ of $%
S_i$ versus $i.$ Here $\tau _i$ is as same as figure 3 and $t_R=\tau
^{\prime }/\tau .$ $\delta \mu $ is the length and $\delta x$ is the maximum
height of period-2 solution in Fig. 3 respectively. The square scatter is
the numerical result and the dotted line is the fitting curves. (a) $\tau
_i=2.026\exp (i/0.817);$ (b) $t_R(i)=1.135\exp (-i/0.747);$ (c) $\delta \mu
(i)=1.072\exp (i/0.616);$ (d) $\delta x(i)=0.125\exp (-i/0.458).$}
\label{fig4}
\end{figure}

\end{document}